%% file: paper.tex
\documentclass[oribibl,pdftex]{llncs}
\usepackage{graphicx}
\usepackage{multirow}
\usepackage{color}
\usepackage{cite}
\usepackage{rotating}
\usepackage{listings}
\usepackage{todonotes}

\newcommand{\Placeholder}[1]{$\langle\!\langle\mbox{\textrm{#1}}\rangle\!\rangle$}

\newcommand{\Src}[1]{\texttt{#1}}

\begin{document}

\title{Parallelization of XPath Queries using \\ Modern XQuery Processors}
\author{Shigeyuki Sato \and Wei Hao \and Kiminori Matsuzaki}
\institute{Kochi University of Technology \\ 
185 Miyanokuchi, Tosayamada, Kami, Kochi, Japan, 782--8502
\email{\{sato.shigeyuki,matsuzaki.kiminori\}@kochi-tech.ac.jp \\
188004h@gs.kochi-tech.ac.jp}
}
\maketitle

\begin{abstract}
\input{abstract}
\end{abstract}

\input{introduction}

\input{approach}

\input{basex}

\input{implementation}

\input{optimization}

\input{experiments}

\input{related}

\input{conclusion}

\paragraph{Acknowledgements}

We would like to thank Christian Gr\"un of the BaseX team for his
technical comments and feedback on BaseX.

\bibliographystyle{splncs04}
\bibliography{reference}
\end{document}

%% file: abstract.tex
A practical and promising approach to parallelizing XPath queries was
proposed by Bordawekar et al. in 2009, which enables parallelization on
top of existing XML database engines.  Although they experimentally
demonstrated the speedup by their approach, their practice has already
been out of date because the software environment has largely changed
with the capability of XQuery processing.  In this work, we
implement their approach in two ways on top of a state-of-the-art XML
database engine and experimentally demonstrate that our implementations
can bring significant speedup on a commodity server.


%% file: introduction.tex
\section{Introduction}\label{sect:introduction}
Scalability and efficiency of XML processing have become increasingly
important because XML data in the real world are steadily
growing up. Parallel XML processing is a natural consequence in the
current hardware environment that allows us to make lavish use of
processor cores. Actually, parallel XQuery processors such as PAXQuery
\cite{camacho-rodriguez15:paxquery} and VXQuery \cite{carman15:vxquery}
were developed for pursuing scalability. They, however, did not
parallelize XPath queries, which characterize a core part of XML
processing. There remains considerable room for improvement particularly
regarding XPath queries.

Parallelization of XPath queries has also been studied but still far
from practical use. Most of the existing studies
\cite{ogden13:pp-transducer,jiang17:ga-pp-transducer,cong12:pe_dist_xpath,nomura07:xpath_tree_skel,kling11:xml_fragment,damigos14:dist_xpath_mapreduce}
were based on divide-and-conquer algorithms on a given XML document.
Although these approaches are algorithmically sophisticated, they are,
unfortunately, impractical in terms of engineering because they confine
input queries to small subsets of XPath and necessitate implementing
dedicated XML database engines. In contrast, Bordawekar et al.\
\cite{bordawekar09:par_xpath} presented an ad hoc yet practical approach
that allows us to use off-the-shelf XML database engines for parallelization
on shared-memory computers; it was based on cheap query split. In this work, we focus on the latter
approach from a practical perspective.

Although Bordawekar et al.'s approach itself is
promising in terms of engineering, their work has already
been out of date in this day and age, particularly in terms of software
environment. A big difference exists in sophistication of
XML database engines. They used an XSLT processor (Xalan-C++ version
1.10) released in 2004, when a matter of the highest priority would have
been standards conformance.
Now, industrial-grade XML database
engines such as BaseX\footnote{\url{http://basex.org/}} are freely
available and we thus can enjoy high-performance queries of high
expressiveness. We therefore
should implement parallelization with serious consideration of
underlying XML database engines, in analogy with existing studies
\cite{grust07:rdbms_xpath,boncz06:monetdb} on efficient XPath query
processing on relational database management systems. In
summary, the following is our research question: \emph{How should we implement
query-split parallelization on state-of-the-art XML database engines?}

To answer this question, in this work, we have developed two
implementations of query-split parallelization on top of BaseX, which
implements XQuery and extensions efficiently \cite{kircher15:xquf_basex}
and involves a powerful query optimizer
\cite{worteler15:func_inline_basex}. We have evaluated our implementations
experimentally with non-trivial queries according to prior work
\cite{bordawekar09:par_xpath} over gigabyte-scale datasets and achieved
significant speedup on a modern server. Through this practice, we have
discovered a simple way of accommodating query-split parallelization to
the query optimization in BaseX.

Our main contributions are summarized as follows.
\begin{itemize}
 \item We have developed two implementations of query-split
       parallelization \cite{bordawekar09:par_xpath} (Sect.\
       \ref{sect:approach}) on top of BaseX (Sect.\ \ref{sect:impl}).
       One is a simple revival of prior work
       \cite{bordawekar09:par_xpath} on BaseX; the other takes advantage
       of rich features of BaseX.
 \item We have investigated the effects of query-split
       parallelization on the query optimization of BaseX and discovered
       a simple way of smooth integration of both (Sect.\ \ref{sect:opt}).
 \item We have experimentally demonstrated that our implementations were
       able to achieve up to 4.86x speedup on a 12-core commodity
       server for non-trivial queries over gigabyte-scale datasets
       (Sect.\ \ref{sect:exper}).
\end{itemize}


%% file: approach.tex
\section{Query-split Parallelization}\label{sect:approach}
In this section, we describe Bordawekar et al.'s approach \cite{bordawekar09:par_xpath} to parallel XPath
query processing. We assume readers' familiarity with
XPath\footnote{\url{https://www.w3.org/TR/xpath/}}.

Bordawekar et al.\ presented a cheap approach to parallel evaluation of
a single XPath query. Their approach is to split a given query into
subqueries in an ad hoc manner and to evaluate them in parallel.
Specifically, they proposed two strategies: query partitioning and data
partitioning.  Query partitioning is to split a given query into
independent queries by using predicates, e.g., from \texttt{$q_1$[$q_2$
or $q_3$]} to \texttt{$q_1$[$q_2$]} and \texttt{$q_1$[$q_3$]}, and from
\texttt{$q_1$/$q_2$} to \texttt{$q_1$[position() <= 10]/$q_2$} and
\texttt{$q_1$[position() > 10]/$q_2$}.  Data partitioning is to split a
given query into a prefix query and a suffix query, e.g., from
\texttt{$q_1$/$q_2$} to prefix $q_1$ and suffix $q_2$, and to run the
suffix query in parallel on each node in the result of the prefix
query. Note that how to construct final results depends on split
queries. If we split and-/or-predicates, we have to intersect/union
results in order. If we perform position-based query partitioning and
data partitioning, we have only to concatenate results in order.

Their approach is a clever engineering choice because it does not
necessitate developing XML processors from scratch and enables us to
extend efficient off-the-shelf XML database engines easily. This is a
great advantage of their approach. A main disadvantage is, in contrast,
that a priori decision of better query split is difficult because of
its ad hoc nature. Particularly, query partitioning depends heavily both
on queries and XML documents. Position-based query partitioning is
almost useless in practice without use of statistics of a given document
as in \cite{bordawekar10:stats_par_xpath}. Data partitioning is easier
to apply and generally outperformed query partitioning in their
experimental comparison \cite{bordawekar09:par_xpath}. In this work, we
therefore deal only with data partitioning.

In this paper, we use the same set of queries over XMark \cite{schmidt02:xmark}
datasets and the DBLP dataset as \cite{bordawekar09:par_xpath} except for DB3, which is a
little modified from the original one for experiments (see
Sect.\ \ref{sect:exper}). Tables~\ref{tbl:query-seq-xmark}--\ref{tbl:query-par-dblp}
summarizes our target queries, where (a)--(c) mean variations of data
partitioning and (a)--(b) except for XM6(b) and DB3(a) follow ones used in
\cite{bordawekar09:par_xpath}.

\input{queries}


%% file: queries.tex
\begin{table}[tbp]
\caption{List of XPath queries for XMark dataset}
\label{tbl:query-seq-xmark}
\footnotesize\centering
\begin{tabular}{l|l}
\hline
 Key & Query \\
\hline
 XM1 & \verb|/site//*[name(.)="emailaddress" or name(.)="annotation"| \\
     & \verb|    or name(.)="description"]| \\
\hline
 XM2 & \verb|/site//incategory[./@category="category52"]/parent::item/@id| \\
\hline
 XM3 & \verb|/site//open_auction/bidder[last()]| \\
\hline
 XM4 & \verb|/site/regions/*/item[./location="United States" and ./quantity > 0| \\
     & \verb|    and ./payment="Creditcard" and ./description and ./name]| \\
\hline
 XM5 & \verb|/site/open_auctions/open_auction/bidder/increase| \\
\hline
 XM6 & \verb|/site/regions/*[name(.)="africa" or name(.)="asia"]/item/| \\
     & \verb|    description/parlist/listitem| \\
\hline
\end{tabular}
\end{table}

\begin{table}[tbp]
\centering
\caption{List of split queries for XMark datasets}
\label{tbl:query-par-xmark}
\footnotesize\centering
\begin{tabular}{l|l}
\hline
 Key & Query \\
\hline
 XM1(a) & prefix = \verb|/site/*|, \\
        & suffix = \verb|descendant-or-self::*[name(.)="emailaddress"| \\
        & \verb|           or name(.)="annotation" or name(.)="description"]| \\
\hline
 XM2(a) & prefix = \verb|/site//incategory|, \\
        & suffix = \verb|self::*[./@category="category52"]/parent::item/@id| \\
 XM2(b) & prefix = \verb|/site/*|, \\
        & suffix = \verb|descendant-or-self::incategory[./@category="category52"]/| \\
        & \verb|           parent::item/@id| \\
 XM2(c) & prefix = \verb|db:attribute("xmark10", "category52")|, \\
        & suffix = \verb|parent::incategory[ancestor::site/parent::document-node()]| \\
        & \verb|           /parent::item/@id| \\
\hline
 XM3(a) & prefix = \verb|/site//open_auction|, \quad suffix = \verb|bidder[last()]| \\
 XM3(b) & prefix = \verb|/site/*|, \\
        & suffix = \verb|descendant-or-self::open_auction/bidder[last()]| \\
 XM3(c) & prefix = \verb|/site/open_auctions/open_auction|, \quad suffix = \verb|bidder[last()]| \\
\hline
 XM4(a) & prefix = \verb|/site/regions/*|, \\
        & suffix = \verb|item[./location="United States" and ./quantity > 0 and | \\
        & \verb|           ./payment="Creditcard" and ./description and ./name]| \\
 XM4(b) & prefix = \verb|/site/regions/*/item|, \\
        & suffix = \verb|self::*[./location="United States" and ./quantity > 0 and| \\
        & \verb|           ./payment="Creditcard" and ./description and ./name]| \\
 XM4(c) & prefix = \verb|db:text("xmark10", "Creditcard")/parent::payment|, \\
        & suffix = \verb|parent::item[parent::*/parent::regions/parent::site/|\\
        & \verb|           parent::document-node()][location = "United States"]| \\
        & \verb|           [0.0 < quantity][description][name]| \\
\hline
 XM5(a) & prefix = \verb|/site/open_auctions/open_auction/bidder|, \quad suffix = \verb|increase| \\
 XM5(b) & prefix = \verb|/site/open_auctions/open_auction|, \quad suffix = \verb|bidder/increase| \\
\hline
 XM6(a) & prefix = \verb|/site/regions/*|, \\
        & suffix = \verb|self::*[name(.)="africa" or name(.)="asia"]/item/| \\
        & \verb|           description/parlist/listitem| \\
 XM6(b) & prefix = \verb|/site/regions/*[name(.)="africa" or name(.)="asia"]/item|, \\
        & suffix = \verb|description/parlist/listitem| \\
\hline
\end{tabular}
\end{table}

\begin{table}[tbp]
\caption{List of XPath queries for DBLP dataset}
\label{tbl:query-seq-dblp}
\footnotesize\centering
\begin{tabular}{l|l}
\hline
 Key & Query \\
\hline
 DB1 & \verb|/dblp/article/author| \\
\hline
 DB2 & \verb|/dblp//title| \\
\hline
 DB3 & \verb|/dblp/book[count(./following-sibling::book[1]/author) | \\
     & \verb|    < count(./author)]| \\
\hline
\end{tabular}
\end{table}

\begin{table}[tbp]
\centering
\caption{List of split queries for DBLP datasets}
\label{tbl:query-par-dblp}
\footnotesize\centering
\begin{tabular}{l|l}
\hline
 Key & Query \\
\hline
 DB1(a) & prefix = \verb|/dblp/article|, \quad suffix = \verb|author| \\
\hline
 DB2(a) & prefix = \verb|/dblp/*|, \quad suffix = \verb|title| \\
 DB2(b) & prefix = \verb|/dblp/*|, \quad suffix = \verb|descendant-or-self::*/title| \\
\hline
 DB3(a) & prefix = \verb|/dblp/book|, \\
        & suffix = \verb|self::*[count(./following-sibling::book[1]/author)| \\
        & \verb|        < count(./author)]| \\
\hline
\end{tabular}
\end{table}

%% file: basex.tex
\section{The XML Database Engine BaseX}\label{sect:basex}
This section briefly describes the XML database engine BaseX. Refer to
the official documentation\footnote{\url{http://docs.basex.org/}} for
more details.

The following are BaseX's features particularly important for this work:
\begin{itemize}
 \item Efficient native implementation of XQuery 3.1, especially
       sequence operations, and XQuery Update Facility
       \cite{kircher15:xquf_basex};
 \item Extension for database operations, especially index-based
       random access;
 \item Extension for text processing, especially
       \Src{ft:tokenize};
 \item Support for in-memory XML databases;
 \item Query optimization based on various indices \cite{worteler15:func_inline_basex};
 \item Support for concurrent transactions in the server mode.
\end{itemize}

The most important (practically essential) feature for our
implementations is the index-based random access to nodes. BaseX offers
indices that enable us to access any node in constant
time. The PRE index, which is an integer that denotes a position in the document order,
brings the fastest constant-time access on BaseX. Function \Src{db:node-pre} returns
the PRE value of a given node and function \Src{db:open-pre} returns the
node of a given PRE value.  PRE values are well suited for representing
the results of prefix queries because only an integer enables us to
restart XPath navigation uniformly from any node.

BaseX implements the whole of XQuery efficiently. Particularly, sequence
operations are very efficient due to sequence implementation based
on finger trees; the length of sequence is returned by
\Src{count} in constant time and the subsequence of a specified range is
extracted in logarithmic time. XQuery Update Facility over in-memory
databases strongly supports efficient use of temporary
databases. Function \Src{ft:tokenize}, which tokenizes a given string to
a sequence of token strings, enables efficient deserialization of
sequences. Our server-side implementation described
in the next section fully utilizes all these features.

BaseX's query optimization is so powerful that it is not unusual to
improve time complexity drastically. For example, the path index enables
pruning traversal of descendants and the attribute index enables instant
access to the nodes that have a specific attribute value. With
aggressive inlining and constant propagation \cite{worteler15:func_inline_basex}, BaseX exploits most constants including
database metadata and PRE values found in a given query for query
optimization. Prevention of spoiling it, as to be described in Sect.\
\ref{sect:opt}, is therefore of crucial importance for performance.

Lastly, BaseX can work efficiently in a client-server model. BaseX
servers can handle concurrent transactions from BaseX clients with
multiple threads. Read transactions including XPath queries are executed
in parallel.


%% file: implementation.tex
\section{Implementing Data Partitioning with BaseX}\label{sect:impl}
In this section, we describe our two implementations called of data
partitioning on top of BaseX. We call them the client-side
implementation and the server-side implementation because of the
difference in the way of managing the results of prefix queries.

Our implementations involve worker threads of simple BaseX clients that
hold independent connections to the BaseX server. After the master
thread issues a prefix query, each worker thread issues a suffix query.

In the rest of this section, we describe our implementations by using
XM3(a) as a running example, assuming the input database to be named
\Src{'xmark'}.  Let $P$ be the number of threads.

\subsection{Client-side Implementation}
The client-side implementation is a simple implementation of data
partitioning with database operations on BaseX.  It sends the server a
prefix query that returns the PRE values of hit nodes as follows.
\begin{lstlisting}
for $x in db:open('xmark')/site//open_auction
 return db:node-pre($x)
\end{lstlisting}

Let this prefix query return sequence \Src{(2, 5, 42, 81, 109, 203)}
through a network. Letting $P = 3$, it is block-partitioned to \Src{(2,
5)}, \Src{(42, 81)}, and \Src{(109, 203)}, each of which is assigned to
a worker thread. To avoid repetitive ping-pong between the client(s) and the server,
we use the following suffix query template:
\begin{lstlisting}[escapechar=\@]
for $x in @\Placeholder{sequence of PRE}@
 return db:open-pre('xmark', $x)/bidder[last()]@\textrm{~,}@
\end{lstlisting}
where \Placeholder{sequence of PRE} is a placeholder to be replaced with a
concrete partition, e.g., \Src{(42, 81)}. Each thread instantiates this
template with its own partition and sends the server the instantiated query.

\subsection{Server-side Implementation}
A necessary task on the results of a prefix query is to block-partition
them. The client-side implementation does it simply on the client side.
In fact, we can also implement it efficiently on the server side by
utilizing BaseX's features.

First, we prepare an in-memory XML database named \Src{'tmp'} initialized
with \Src{<root> </root>}, which is a temporary database for storing the
results of a prefix query. The prefix query is implemented as follows:
\begin{lstlisting}[escapechar=\@]
let $P := @\Placeholder{number of partitions}@
let $s := db:open('xmark')/site//open_auction ! db:node-pre(.)
for $i in 1 to $P
 return insert node element part { $blk_part($i, $P, $s) }
         as last into db:open('tmp')/root@\textrm{~,}@
\end{lstlisting}
where \Placeholder{number of partitions} denotes a placeholder to be
replaced with a concrete value of $P$ and \verb|$blk_part($i, $P, $s)|
denotes the \verb|$i|th one of \verb|$P| partitions of \verb|$s|
implemented in logarithmic time with sequence operations.

In the example case used earlier, \Src{'tmp'} database results in the
following:
\begin{lstlisting}[escapechar=\@]
@$^1$@<root>
 @$^2$@<part>@$^3$@2 5</part>@$^4$@<part>@$^5$@42 81</part>@$^6$@<part>@$^7$@109 203</part>
@$^{\phantom{1}}$@</root>@\textrm{~,}@
\end{lstlisting}
where a left superscript denotes a PRE value. Note that its document
structure determines the PRE value of $i$th partition to be $2i+1$.

A suffix query is implemented with deserialization of a partition as
follows:
\begin{lstlisting}[escapechar=\@]
for $x in ft:tokenize(db:open-pre('tmp', @\Placeholder{PRE of partition}@))
 return db:open-pre('xmark', xs:integer($x))/bidder[last()])@\textrm{~,}@
\end{lstlisting}
where \Placeholder{PRE of partition} denotes a placeholder to be replaced
with the PRE value of a target partition and the care of empty partitions
is omitted for brevity.

The server-side implementation is more efficient because transferred
data between clients and a server except for output are in a
constant size.


%% file: optimization.tex
\section{Integration with Query Optimization}\label{sect:opt}
As mentioned in Sect.\ \ref{sect:basex}, BaseX is equipped with a powerful
query optimizer. For example, BaseX optimizes XM3 to
\begin{lstlisting}
/site/open_auctions/open_auction/bidder[last()]
\end{lstlisting}
on the basis of the path index, which brings knowledge that
\Src{open\_auction} exists only immediately below \Src{open\_auctions}
and \Src{open\_auctions} exists only immediately below
\Src{site}. The search space of this optimized query has significantly
reduced because an expensive step of descendant-or-self axis is replaced
with two cheap steps of child axes. It is worth noting that a more
drastic result is observed in
XM2, where the attribute index is exploited through function
\Src{db:attribute}.

Data partitioning converts a given query to two separate ones and
therefore affects the capability of BaseX in query optimization. In
fact, the suffix query of XM3(b) is not optimized to the
corresponding part of optimized XM3 because BaseX does not utilize
indices for optimizing queries starting from nodes specified with PRE
values even if possible in principle. Most index-based optimizations are
limited to queries starting from the document root.  This is a
reasonable design choice in query optimization because it is expensive
to check all PRE values observed. However, it is unnecessary to check any
PRE value that specifies the starting nodes of the suffix query because
of the nature of data partitioning, of which BaseX is unaware.  This
discord between BaseX's query optimization and data partitioning may incur
serious performance degradation.

A simple way of resolving this discord is to apply data partitioning
after BaseX's query optimization. Data partitioning is applicable to any
multi-step XPath query in principle. Even if an optimized query is
thoroughly different from its original query as in XM2, it is
entirely adequate to apply data partitioning to the optimized query,
forgetting the original. In fact, XM2--4(c) are instances of such
data partitioning after optimization.
This coordination is so simple that we are still able to implement data
partitioning only by using BaseX's dumps of optimized queries without
any modification on BaseX. This is a big benefit.


%% file: experiments.tex
\section{Experiments}\label{sect:exper}
In this section, we describe the experimental evaluation of our
implementations.

\subsection{Experimental Setting}
We have conducted several experiments to evaluate the performance
of the two implementations of parallel XPath queries.
All the experiments were conducted on a computer that equipped
with two Intel Xeon E5-2620 v3 CPUs ($2 \times 6$ cores, 2.4GHz, Hyper-Threading off)
and 32-GB memory (DDR4-1866). The software environment we used was Java
OpenJDK 64-Bit Server VM (ver. 9-internal, \texttt{9\~{}b114-0ubuntu1} in Ubuntu
16.04 LTS) and BaseX ver. 9.0.1 (\texttt{3a8b2ad6}) with minor
modifications to enable TCP\_NODELAY.

We used two datasets: XMark and DBLP. We generated an XMark dataset with XMLgen\footnote{\url{https://projects.cwi.nl/xmark/downloads.html}} 
giving \Src{-f 10}, which was of 1.1 GB and had 16 million nodes.
The root of the XMark tree has six children \verb|regions|, \verb|people|,
\verb|open_auctions|, \verb|closed_auctions|, \verb|catgraph|, and \verb|categories|,
which have 6, 255000, 120000, 97500, 10000, and 10000 children, respectively.
Refer to~\cite{schmidt02:xmark} for more details of the XMark dataset.
The DBLP dataset was the latest one downloaded from the DBLP
website\footnote{\url{https://dblp.uni-trier.de/xml/}}, where the date
of the downloaded file was August 29, 2017 and the dataset was of 2.2 GB
and had 53 million nodes. The DBLP tree was flat; the root element has 6
million children.

We used the XPath queries shown in
Tables~\ref{tbl:query-seq-xmark}--\ref{tbl:query-par-dblp}, which are
the same as those used in \cite{bordawekar09:par_xpath} except for
DB3. We modified DB3 to alleviate the computational cost because
the original one costs quadratic time, which was too costly to run over
the latest DBLP dataset. We measured execution time until the client
received all the results of suffix queries into byte streams.
We executed both the client-side implementation and the server-side one
for each parallel XPath query.
The execution time does not include the loading time, that is,
the input dataset was loaded into memory before the execution of queries.
We measured the execution time of 25 runs after a warm-up run and
picked up their median.

\subsection{Total Execution Time}
\input{exp_results/summary}

Table~\ref{Table:summary} summarizes the execution time of the queries.
The ``orig $t_o$''column shows the time for executing
original queries XM1--XM6 and DB1--DB3 with BaseX's \verb|xquery| command.
The ``seq $t_s$'' columns show the time for executing
the prefix query and the suffix query with one thread.
The ``par $t_p$'' columns show the time for executing
the prefix query with one thread and the suffix query with 12 threads.
The table also includes for reference
the speedup of parallel queries with respect to original queries
and the size of results of the prefix queries and the whole queries.

By using the pair of the prefix and suffix queries split at an appropriate
step, we achieved speedups by factor about 2.0 for XM1 and XM3,
and by factor of more than 3.7 for XM4 and XM6.
The execution time of XM2 was very short because
BaseX executed an optimized query that utilized the attribute index as mentioned in Sect.\ \ref{sect:opt}.
By designing the parallel query XM2(c) based on that optimized query,
the execution time of parallel query was just longer than that of
the optimized query by 2 ms.
Comparing the two implementations,
we observed that the server-side implementation ran faster for most queries.

Although some of parallel queries did not reached the performance of
their original queries, these were reasonable.  XM2(a)--(b) were due to
index-based optimizations; XM2(c) was too cheap to benefit from parallel
evaluation. XM4(a) and XM6(a) were due to load imbalance derived from
data skewness. XM5(a) and DB3(a) were due to the cheapness of the suffix
queries compared to the prefix ones; their suffix queries visited few
nodes from a starting one and merely filtered the results of their
prefix queries.

\subsection{Breakdown of Execution Time}
\input{exp_results/breakdown}

To investigate the execution time in detail,
we executed parallel queries XM1(a), XM3(c), XM4(c), XM5(b), XM6(b),
DB1(a), DB2(a) and DB3(a)
with $P=1$, 2, 3, 6, and 12 threads.
Tables~\ref{table:breakdown-c} and \ref{table:breakdown-s} show
the breakdown of the execution time divided into two phases: prefix query and suffix query.
In these tables, the speedup is calculated with respect to the
execution time of suffix queries with one thread.

From Tables~\ref{table:breakdown-c} and \ref{table:breakdown-s}, we can
find several interesting observations. First, the execution time of
prefix queries was almost proportional to their result sizes and almost
the same between the two implementations (except for DB2(a)). Comparing the two
implementations, we can observe that the server-side implementation
performed better than the client-side implementation
for most suffix queries.  These results suffice for
concluding that the server-side implementation is, as expected, more
efficient.

Next, we analyze the dominant factor of the performance gaps between the
client-side and the server-side. Although the performance gaps of prefix
queries should be mainly the difference between sending data to clients
on localhost and storing data into memory, it was not significant.
Communication cost, which is our expected advantage of the server-side,
therefore did not explain the dominant factor of total performance gaps.

By examining the logs of the BaseX server, we have found that the
dominant factor was the parsing of suffix queries. Since the client-side
implementation sends a suffix query of length linearly proportional to the
result size of a prefix query, it can be long. In fact, the prefix query
of DB1(a) resulted in 15.8 MB and BaseX took more than 700 ms for
parsing the query string in the client-side implementation; it took less than 1 ms in the server-side implementation.  Note that sending and receiving a long query would not cost so much in our experiments because localhost communication was almost as fast as local memory
access.  Parsing in the client-side implementation is more expensive
than deserialization in the server-side implementation.
More importantly, parsing is essentially not streamable: before finishing parsing a
query string, BaseX cannot start to evaluate it, whereas the deserialization in
the server-side is streamable. We conclude that this difference in
streamability was the dominant factor of the performance gaps between
the client-side and the server-side.

Lastly, we should admit a conundrum that we could not solve. Several
queries performed poorly at specific $P$; e.g., DB1(a) at $P=3$ for the
client-side implementation and XM4(c) and XM5(b) at $P=6$ for the
server-side implementation resulted unreasonably. Because these results
were reproducible, we conjecture that they were derived from the mechanism
of BaseX.

\subsection{Scalability Analysis}
When we analyze the speedup of parallel execution,
the ratio of sequential execution part to the whole computation is important
because it limits the possible speedup by Amdahl's law.
In the two implementations, most of the sequential execution part consists in the prefix query. The ratio of the sequential execution part was small in general:
more specifically,
the client-side implementation had smaller ratio (e.g. 12.6 \% for DB1(a)) than
the server-side implementation had (e.g. 16.6 \% for DB1(a)).
In our implementations, the suffix queries were independently executed in parallel
through individual connections to the BaseX server.
The speedups we observed for the suffix queries were, however, smaller than we
had expected.  We also noticed that in some cases the execution time was longer
with more threads (for example, XM3(c) P=12 with the server-side implementation).

\input{exp_results/load-work}

To understand the reason why the speedups of the suffix queries were small,
we made two more analyses (Table~\ref{tbl:load-work}).
The degree of load balance in processing suffix queries was
calculated as the cumulative execution time divided the maximum execution time:
\[
 \mbox{degree of load balance} = \frac{\sum t_i^{p}}{\max t_i^p}
\]
where $t_i^p$ denotes the execution time of the $i$th suffix query in
parallel with $p$ threads.
The increase of work of the suffix queries was
calculated by the cumulative execution time divided by the single-thread execution time:
\[
 \mbox{degree of increase of work} = \frac{\sum t_i^{p}}{t_1^{1}}~.
\]

From Table~\ref{tbl:load-work}, we can observe
the reasons of the small speedups in the suffix queries.
Obvious load imbalances were incurred in XM1(a) and DB3(a) for different
reasons. For XM1(a), the hit nodes by the prefix query were very few and
less than the number of cores. For DB3(a), the computational cost of
each suffix query was quite different because of data skewness.
For the other cases, we achieved good load balance,
and the degrees of load balance were more than 75\% even with 12 threads,
which means that load imbalance was not the main cause of small speedups for those
queries.
The increase of work was significant for XM5(b) and XM3(c),
and it was the main cause that the queries XM5(b) and XM3(c) had small speedups.
For the other queries, we observed very small increase of work until 6 threads,
but the work increased when 12 threads.
Such an increase of work is often caused by contention to memory access,
and it is inevitable in shared-memory multicore computers.


%% file: exp_results/summary.tex
\begin{table}[tbp]
\caption{Summary of execution time.}
\label{Table:summary}
\setlength{\doublerulesep}{.4pt}
\centering\begin{tabular}{l|r|rr@{~(}r@{)}|rr@{~(}r@{)}|rr}
\hline \hline
~~Key & orig $t_o$~ & \multicolumn{3}{c|}{client-side} & \multicolumn{3}{c|}{server-side} & \multicolumn{2}{c}{Result size} \\
	&       		& seq $t_s$~   	& par $t_p$~   	& $t_o/t_p$	& seq $t_s$~	& par $t_p$~	& $t_o/t_p$	& prefix~	& final~ \\
\hline
XM1(a)	& 37263			& 44443		& 18058		& 2.06		& 40084		& 16137		& 2.31		& 54 B		& 994 MB \\
\hline
XM2(a)	& \multirow{3}{*}{2}	& 2856		& 1055		& 0.00		& 1075		& 808		& 0.00		& 6.62 MB	& \multirow{3}{*}{1.55 KB} \\
XM2(b)	&			& 1029		& 937		& 0.00		& 1049		& 902		& 0.00		& 54 B		&  \\
XM2(c)	&			& 3		& 4		& 0.50		& 3		& 4		& 0.50		& 671 B		&  \\
\hline
XM3(a)	& \multirow{3}{*}{639}	& 1180		& 304		& 2.10		& 848		& 302		& 2.12		& 1.08 MB	& \multirow{3}{*}{14.5 MB} \\
XM3(b)	&			& 1816		& 1663		& 0.38		& 1857		& 1490		& 0.43		& 54 B		&  \\
XM3(c)	&			& 1154		& 305		& 2.10		& 850		& 321		& 1.99		& 1.08 MB	&  \\
\hline
XM4(a)	& \multirow{3}{*}{1148}	& 1595		& 1595		& 0.72		& 1647		& 1084		& 1.06		& 49 B		& \multirow{3}{*}{26.4 MB} \\
XM4(b)	&			& 1858		& 545		& 2.11		& 1402		& 493		& 2.33		& 1.75 MB	&  \\
XM4(c)	&			& 1121		& 245		& 4.69		& 1232		& 236		& 4.86		& 106 KB	&  \\
\hline
XM5(a)	& \multirow{2}{*}{715}	& 2535		& 828		& 0.86		& 1288		& 651		& 1.10		& 5.38 MB	& \multirow{2}{*}{15.9 MB} \\
XM5(b)	&			& 1209		& 462		& 1.55		& 955		& 432		& 1.66		& 1.08 MB	&  \\
\hline
XM6(a)	& \multirow{2}{*}{820}	& 954		& 929		& 0.88		& 970		& 933		& 0.88		& 49 B		& \multirow{2}{*}{22.2 MB} \\
XM6(b)	&			& 1004		& 219		& 3.74		& 1084		& 207		& 3.96		& 183 KB	&  \\
\hline
DB1(a)	& 6759                  & 12498		& 4185		& 1.62		& 8451		& 2730		& 2.48		& 15.8 MB	& 176 MB \\
\hline
DB2(a)	& \multirow{2}{*}{15641}& 34729		& 8555		& 1.83		& 19191		& 6082		& 2.57		& 56.9 MB	& \multirow{2}{*}{423 MB} \\
DB2(b)	&                       & 34713		& 8564		& 1.83		& 19105		& 6405		& 2.44		& 56.9 MB	& \\
\hline
DB3(a)	& 888                   & 1115		& 1092		& 0.81		& 1043		& 953		& 0.93		& 139 KB	& 1.9 MB \\
\hline
\end{tabular}
\end{table}

%% file: exp_results/breakdown.tex
\begin{table}[tbp]
\caption{Breakdown of execution time for client-side implementation}
\label{table:breakdown-c}
\setlength{\doublerulesep}{.4pt}
\centering\begin{tabular}{l|r|rrrrr@{~~(}c@{)}}
\hline
\hline
        & prefix        & \multicolumn{6}{c}{suffix \raisebox{-.7pt}{$t^P$}}                                                   		\\
	& 		& \makebox[4.2em][c]{P=1}	& \makebox[4.2em][c]{P=2}	& \makebox[4.2em][c]{P=3}	& \makebox[4.2em][c]{P=6}	& \makebox[4.2em][c]{P=12} & $t^{1}/t^{12}$ \\
\hline
XM1(a)	& 5		& 44438		& 27063		& 27296		& 27601		& 18053		& 2.46		 \\
XM3(c)	& 93		& 1061		& 656		& 423		& 251		& 212		& 5.00		 \\
XM4(c)	& 23		& 1098		& 766		& 563		& 310		& 222		& 4.95		 \\
XM5(b)	& 89		& 1120		& 716		& 549		& 395		& 373		& 3.00		 \\
XM6(b)	& 21		& 983		& 691		& 525		& 268		& 198		& 4.96		 \\
\hline
DB1(a)	& 1395		& 11103		& 6172		& 10992		& 5614		& 2790 		& 3.98		 \\
DB2(a)	& 2748		& 31981		& 20502		& 20077		& 9401		& 5807		& 5.51		 \\
DB3(a)	& 602		& 513		& 562		& 555		& 527		& 490		& 1.05		 \\
\hline
\end{tabular}
\end{table}

\begin{table}[tbp]
\caption{Breakdown of execution time for server-side implementation}
\label{table:breakdown-s}
\setlength{\doublerulesep}{.4pt}
\centering\begin{tabular}{l|r|rrrrr@{~~(}c@{)}}
\hline
\hline
        & prefix        & \multicolumn{6}{c}{suffix \raisebox{-.7pt}{$t^P$}}                                                   		\\
	& 		& \makebox[4.2em][c]{P=1}	& \makebox[4.2em][c]{P=2}	& \makebox[4.2em][c]{P=3}	& \makebox[4.2em][c]{P=6}	& \makebox[4.2em][c]{P=12} & $t^{1}/t^{12}$ \\
\hline
XM1(a)	& 7		& 40077		& 27162		& 26308		& 16342		& 16130		& 2.48		 \\
XM3(c)	& 78		& 772		& 446		& 322		& 209		& 243		& 3.18		 \\
XM4(c)	& 18		& 1214		& 736		& 548		& 1668		& 218		& 5.57		 \\
XM5(b)	& 78		& 877		& 545		& 412		& 691		& 354		& 2.48		 \\
XM6(b)	& 20		& 1064		& 632		& 482		& 257		& 187		& 5.69		 \\
\hline
DB1(a)	& 1204		& 7247		& 4165		& 2861		& 1584		& 1526 		& 4.75		 \\
DB2(a)	& 2251		& 16940		& 11470		& 7962		& 4520		& 3831		& 4.42		 \\
DB3(a)	& 518		& 525		& 439		& 450		& 415		& 435		& 1.21		 \\
\hline
\end{tabular}
\end{table}

%% file: exp_results/load-work.tex
\begin{table}[tbp]
 \caption{Load balance and increase of work of server-side implementation (load-balance/increase-of-work)}
 \label{tbl:load-work}
  \centering
  \begin{tabular}{l|r@{/}rr@{/}rr@{/}rr@{/}r}
\hline
	& \multicolumn{2}{c}{$P=2$}& \multicolumn{2}{c}{$P=3$}& \multicolumn{2}{c}{$P=6$}& \multicolumn{2}{c}{$P=12$} \\
\hline
XM1(a)	& ~1.51	& 1.02	& ~~1.59	& 1.04	& ~~2.83	& 1.15	& ~~~2.87	& 1.15 \\
XM3(c)	& 1.95	& 1.12	& 2.84	& 1.18	& 5.38	& 1.46	& 9.01	& 2.84 \\
XM4(c)	& 1.97	& 1.19	& 2.82	& 1.27	& 2.59	& 3.55	& 10.25	& 1.84 \\
XM5(b)	& 1.91	& 1.19	& 2.79	& 1.31	& 3.97	& 3.13	& 11.31	& 4.57 \\
XM6(b)	& 1.94	& 1.15	& 2.76	& 1.25	& 5.40	& 1.30	& 10.35	& 1.82 \\
\hline
DB1(a)	& 1.95	& 1.12	& 2.86	& 1.13	& 5.69	& 1.24	& 10.79	& 2.27 \\
DB2(a)	& 1.58	& 1.07	& 2.38	& 1.12	& 4.68	& 1.25	& 9.80	& 2.22 \\
DB3(a)	& 1.16	& 0.97	& 1.24	& 1.06	& 1.45	& 1.15	& 1.83	& 1.52 \\
\hline
  \end{tabular}
\end{table}

%% file: related.tex
\section{Related Work}\label{sect:related}
Most of existing studies
\cite{ogden13:pp-transducer,jiang17:ga-pp-transducer,cong12:pe_dist_xpath,nomura07:xpath_tree_skel,kling11:xml_fragment,damigos14:dist_xpath_mapreduce,hao16:partial_tree,bi17:twig_mr}
on parallel evaluation of XPath queries were based on divide-and-conquer
algorithms on a given document. They adopted different ways of dividing a
 document into fragments. Cong et al. \cite{cong12:pe_dist_xpath} and Nomura
et al. \cite{nomura07:xpath_tree_skel} adopted a tree-shaped fragment
that contains original nodes and hole nodes, where a hole node
represents a link to a missing subtree, and represented the whole
document as a tree of fragments. The key part of their approaches to
decouple dependencies between evaluations on fragments so as to perform
them in parallel. Kling et al. \cite{kling11:xml_fragment} modeled
fragmentation as horizontal and vertical in terms of schemas. Horizontal
fragmentation is to divide a document tree with replication of top part
so as to preserve the path from the root to each node of every fragment,
i.e., preserve its schema, where the whole document is a simple
collection of fragments. Vertical fragmentation is to divide, in
contrast, a document tree together with its schema, where each fragment
is a forest conforming a fragmented schema, and thus to represent the
whole document as connected graphs of fragments. In parallel pushdown
transducers \cite{ogden13:pp-transducer,jiang17:ga-pp-transducer}, a
given document is modeled as a sequence of matched brackets and a
fragment is represented as a sequence of unmatched brackets, which are,
in contrast, modeled as a special kind of tree nodes in partial trees
\cite{hao16:partial_tree}.

All these approaches are quite advantageous to scalability. It has been
exemplified by their accommodation
\cite{cong12:pe_dist_xpath,damigos14:dist_xpath_mapreduce,bi17:twig_mr} to processing by
Hadoop MapReduce, which is a
common infrastructure for large-scale data processing, and to parallel
streaming \cite{ogden13:pp-transducer,jiang17:ga-pp-transducer}.
Unfortunately, they are inconvenient in practice because their
algorithms were designed for evaluating small subsets (e.g., child and
descendant axes with predicates in
\cite{cong12:pe_dist_xpath,kling11:xml_fragment}) of XPath often by using
dedicated data structures for XML data storage. The necessity of such
internal data structures imposes implementing XML database engines
nearly from scratch in practice. Even a fallback into serial evaluation
of queries beyond a tractable subset then becomes nontrivial. In
summary, these approaches are hard in engineering.

Surprisingly, Bordawekar et al.'s approach \cite{bordawekar09:par_xpath}
has not been well studied regardless of its great virtue in
engineering. They studied by themselves on the sophistication of the
query split strategies based on the statistics of a given document
\cite{bordawekar10:stats_par_xpath}. Since that, follow-up studies had
not been seen for a long time, but most recently, Karsin et al.\
\cite{karsin17:sched_xpath_par} has studied on the scheduling of suffix
query tasks. They investigated three kinds of task generation and showed
experimentally their trade-off between overhead and load imbalance.
However, they did not focus on the great advantage in engineering and
used their own sequential query engine for pursuing experimental
performance predictability. In contrast, our work focuses on how to
integrate Bordawekar et al.'s approach into state-of-the-art XML
database engines from a practical perspective.


%% file: conclusion.tex
\section{Conclusion}\label{sect:concl}
In this paper, we have reassesed data-partitioning parallelization of
XPath queries proposed by Bordawekar et al.\
\cite{bordawekar09:par_xpath} on top of BaseX. We have developed two
implementations on the basis of BaseX's features. The server-side
implementation, which particularly exploits rich features of BaseX,
achieved roughly better experimental speedup because of lower overhead
in querying.

Although this paper focuses only on BaseX, we have also tried to
implement data partitioning on top of other XML processors
(e.g., Xalan-Java) and XML database management systems (e.g., Microsoft SQL
Server).  Unfortunately, implementing data partitioning on top of
Xalan-Java and SQL Server has turned to be practically infeasible. The
latest implementation of Xalan-Java is designed for stream processing
and therefore provides no API for index-based access. Although SQL
Server 2017 provides an API for XML indices, direct manipulation of
their actual values is prohibited and operations are
quite restrictive. In contrast, BaseX allows users to manipulate index
values directly and provides various efficient operations on them. We therefore
have made a success of implementing data partitioning in two
different ways. That is, our work also has the implication of
demonstrating practical requirements on XML database engines for data
partitioning---we believe it to be helpful for development of XML database systems.
